\input{aipcheck}

\documentclass[
    ,final            % use final for the camera ready runs
  ]
  {aipproc}

\layoutstyle{8x11double}

\begin{document}

\title{High Resolution He-like Argon And Sulfur Spectra From The PSI ECRIT}

\author{ \underline {M.~Trassinelli}}{
  address={Laboratoire Kastler Brossel, Universit\'e Pierre et Marie Curie, Paris, France}
}

\author{S.~Biri}{
  address={Insitute of Nuclear Research (ATOMKI), Debrecen, Hungary}
}

\author{S.~Boucard}{
  address={Laboratoire Kastler Brossel, Universit\'e Pierre et Marie Curie, Paris, France}
}

\author{D.S.~Covita}{
  address={Physics Department, University of Coimbra, Portugal}
}

\author{D.~Gotta}{
  address={Institut f\"ur Kernphysik, Forschungszentrum J\"ulich, J\"ulich, Germany}
}

\author{B.~Leoni}{
  address={Paul Scherrer Institut, Villigen PSI, Switzerland}
}

\author{A.~Hirtl}{
  address={\"Osterreichisch Akademie der Wissenschaften, Wien, Austria}
}

\author{P.~Indelicato}{
  address={Laboratoire Kastler Brossel, Universit\'e Pierre et Marie Curie, Paris, France}
}

\author{E.-O.~Le~Bigot}{
  address={Laboratoire Kastler Brossel, Universit\'e Pierre et Marie Curie,  Paris, France}
}

\author{J.M.F.~dos Santos}{
  address={Physics Department, University of Coimbra, Portugal}
}

\author{L.M.~Simons}{
  address={Paul Scherrer Institut, Villigen PSI, Switzerland}
}

\author{L.~Stingelin}{
  address={Paul Scherrer Institut, Villigen PSI, Switzerland}
}

\author{J.F.C.A.~Veloso}{
  address={Physics Department, University of Coimbra}
  ,altaddress={Physics Department, University of Aveiro, Portugal}
}
\author{A.~Wasser}{
  address={Paul Scherrer Institut, Villigen PSI, Switzerland}
}

\author{J.~Zmeskal}{
  address={\"Osterreichisch Akademie der Wissenschaften, Wien, Austria}
}

\begin{abstract}
We present new results on the X-ray spectroscopy of multicharged argon, sulfur and chlorine obtained
with the Electron Cyclotron Resonance Ion Trap (ECRIT) in operation at the Paul Scherrer Institut 
(Villigen, Switzerland). We used a Johann-type Bragg spectrometer with a spherically-bent 
crystal, with an energy resolution of about 0.4 eV\@. The ECRIT itself is 
of a hybrid type, with a superconducting split coil magnet, special iron 
inserts which provides the mirror field, and a permanent magnetic hexapole. The high 
frequency was provided by a 6.4 GHz microwave emitter.

We obtained high intensity X-ray spectra of multicharged F-like to He-like argon, sulfur and chlorine
 with one 1s hole. In particular, we observed the $1s2s\;^{3}S_1 \to 1s^2\;^{1}S_0\; M1$ and 
$1s2p\;^{3}P_2 \to 1s^2\;^{1}S_0\; M2$  transitions in He-like argon, sulfur and chlorine with unprecedented 
statistics and resolution. The energies of the observed lines are being determined with good 
accuracy using the He-like M1 line as a reference.

%Moreover we scanned the plasma by focusing the spectrometer to different points in 
%the plasma chamber. 
We surveyed the He-like M1 transition intensity as a function 
of the ECRIT working conditions. In particular we observed the M1 intensity 
dependency on the coil current and on the injected microwave power. 
\end{abstract}

\maketitle

%%%%%%%%%%%%%%%%%%%%%%%%%%%%%%%%%%%%%%%%%%%%
%% MAINMATTER
%%%%%%%%%%%%%%%%%%%%%%%%%%%%%%%%%%%%%%%%%%%%

\section{Introduction}
The Electron Cyclotron Resonance Ion Trap (ECRIT) of the Paul Scherrer Institut (PSI) has been
set up with the goal of producing narrow X-ray lines that would yield 
the response function of the Bragg crystal spectrometer with high accuracy~\cite{Gotta2004}. Such a measurement is crucial for 
the  ongoing pionic hydrogen experiment at PSI~\cite{Biri2000,Simons2003,proposal}. 
For low- to medium-Z atoms, the E1 X-rays from hydrogen-like ions and the M1 X-rays from helium-like 
ions have an energy of a few keV, with a natural width which is negligible compared to 
the expected resolution of the Bragg crystal spectrometer. 
The ion kinetic energy  in an ECRIS is small: it is on the level of less than 1~eV~\cite{Bernard1996,Sadeghi1991}. 
Because of this, a Doppler broadening of less than 40~meV can 
be expected in transitions in the 3~keV energy range in He-like argon.

In 2002, the $1s2s\;^{3}S_1 \to 1s^2\;^{1}S_0\; M1$ X-ray transition in He-like argon  has been 
used to study the response function of 
the Bragg crystal spectrometer with quartz [10-1] and silicon [111] crystals~\cite{Simons2004}. 
In spring 2004, the characterization of the spectrometer was completed with the test of our
whole set of crystals (quartz [10-1], quartz [100] and silicon [111]) using the M1 transition line 
from He-like argon, chlorine and sulfur.
In addition, we obtained  X-ray spectra of highly-charged ions of these elements with unprecedented 
statistics and resolution.

\section{Experimental Set-up}
\begin{figure}[h]

\includegraphics[width=.475\textwidth]{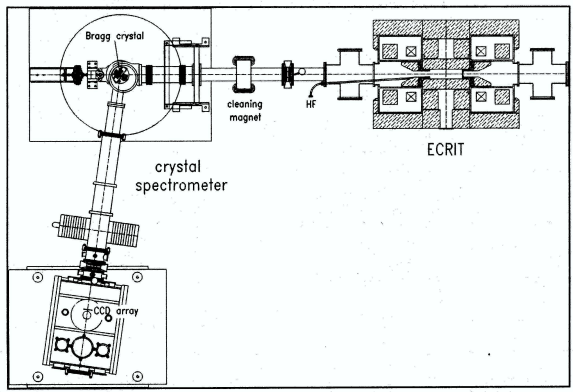} \label{setup}
  \caption{Set-up of the experiment.}
\end{figure}

The experimental set-up is almost identical to the 2002 run set-up described in Ref.~\cite{Simons2004}.
It is composed of the ECRIT source and of a Bragg crystal 
spectrometer coupled to a position sensitive detector (see fig.~\ref{setup}). 

%\subsubsection{The ECRIT Source}
The ECRIT consists of a pair of a superconducting 
split coil magnets (which, together with special iron inserts, provides the mirror field 
configuration), of an AECR-U style permanent hexapole magnet, and of a 6.4~GHz power regulated
emitter. The mirror field parameters provide one of the highest mirror ratios for ECR
sources, with a value of 4.3 over the length of the plasma chamber. 
%The exapole is 
%cooled at the front pieces and on the inner radius by a forced flow of demineralised 
%water. The plasma chamber is formed by a 0.4~mm thick stainless steel tube of inner 
%diameter of 85~mm and a length of 265~mm axially limited by copper inserts.  
%At the position of the exapole gap the stainless steel tube is perforated by a series 
%of diameter 2.5~mm holes allowing for radial pumping in additional to the axial 
%pumping. 
%The pumping system was composed by three 3000~l/min turbolecular pumps and a 1000~l/s cryo-pump. 

By using a cryopump and reducing the surface of iron insertion pieces 
the reference pressure ( i.e. without plasma) was reduced from $1.7 \cdot 10^{-7}$~mbar (2002)
to $3 \cdot 10^{-8}$~mbar
%A reference pressure (without plasma) 
%of $3 \cdot 10^{-8}$~mbar ($6 \cdot 10^{-8}$~mbar with 
%the cryo-pump off)
%was reached, after reducing the surface of the iron insertion pieces from the 2002 
%set-up and installing a cry-pump; this pressure was $1.7 \cdot 10^{-7}$~mbar in 2002.
Gas filling was supplied radially through the gaps in the open structure hexapole. The 
gas composition was routinely controlled and stabilized  with a quadrupole mass spectrometer.

%\begin{figure}[t!]
%\includegraphics[width=0.8\textwidth]{ecr32_notxt}
%  \caption{Schematic of the PSI ECRIT, with axial field values superimposed. 
%1) Iron return yoke and force balance iron. 2) Exapole. 3) Field forming iron.
% 4) Superconducting coils. 5) high frequency (HF) wave guides. 6) Copper inserts. }
%\end{figure}

%\subsubsection{Crystal Spectrometer and CCD detector}
A Bragg  crystal spectrometer of reflection type (Johann configuration) was installed 
at a distance of 2200~mm from the plasma center. Silicon and quartz crystal have
used for the 2004 run, with different Bragg angle values. The crystals are 
100~mm diameter circular plates, with a thickness of 0.2--0.3~mm, and are
spherically bent  with a curvature radius of $2982.4 \pm 0.6$~mm, 
by optical attachment to high-quality quartz spherical lenses.

The detector is an array of 6 CCDs of $600 \times 600$
pixels each~\cite{Nelms2002}, with an energy resolution of 140~eV at 3~keV. 
The pixel size at working temperature ($-100^\circ C$) has 
been recently measured to be $39.9943 \pm 0.0035~\mu m$~\cite{Trassinelli2004}. 
The granularity of the detector was decisive in discriminating charged particle 
events against X-rays possessing different topologies. The CCD chips and the associated
electronics were  protected against light and high-frequency (HF) power by a 
$30~\mu m$ thick Beryllium 
window installed in the vacuum tube between the crystal spectrometer and the ECRIT.

\subsection{ECRIT at work}
During spring 2004, we injected different kinds of gases in the ECRIT in order to obtain X-ray spectra from highly-charged argon, chlorine and sulfur. For this propose, we used a gas mixture of $O_2$
and, respectively, $Ar$, $CHClF_2$ and $SO_2$. Using the experience acquired in the 
previous run, we adjusted a mixing ratio of around 1:9, with a total pressure in the plasma 
chamber of  $3-4 \cdot 10^{-7}$~mbar. 
In order to recognize the different charge states, we used as an initial  reference the 
$K \alpha$ or $K \beta$ lines of the neutral gas, which are easily recognizable: they are the brightest when only a few watts of HF power 
are injected. Using known energy intervals, we were then able to move the spectrometer to the region of the nearby 
$1s2s\;^{3}S_1 \to 1s^2\;^{1}S_0\; M1$ transition in the He-like ion, and to observe it.
We then optimized the different ECRIT and spectrometer parameters in order to maximize the line intensity 
in the detector.
The typical illumination time of the CCD chips before the readout is 1~min. Due to the large 
intensity of the X-rays, this allows
 for a sizeable probability of double hits for pixels near the line peak. 
In order to reduce this effect and improve the peak-to-background ratio, a densimet\textregistered{} collimator
was inserted at a distance of 150~mm from the center of the plasma, leaving an aperture
of 28~mm(h) $\times$ 4~mm(v) or 28~mm(h) $\times$ 1~mm(v), 
depending on the configuration.
During the different runs  the (double hits) to (single hits) ratio always stayed below 5\%, 
which is small enough 
to properly handle double hits in the final analysis, and to neglect triple hit processes.

%\subsection{Results}

%During the optimization of the system, we could observe the dependency of the M1 line 
%intensity vs. the superconducting coils current, vs. the gas pressure and ratio and 
%vs. the HF injected power.
%As mention in the precedent paragraph, during the pressure scan we found the optimal 
%values  with mixing ratio around 1:9 with a total pressure in the plasma 
%chamber of  $3-4 \cdot 10^{-7}$~mbar. 
%The formation of the He-like depends crucially on the injected HF power with a marked 
%threshold effect. 

\begin{figure}[h]
\includegraphics[height=.3\textheight,]{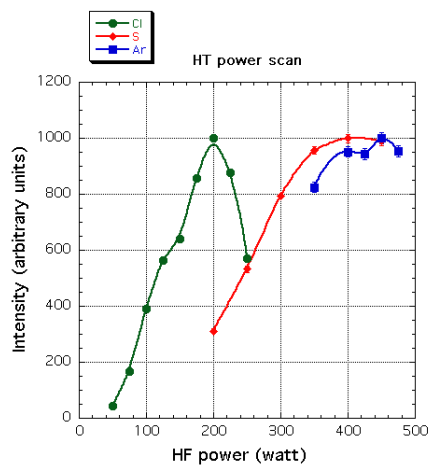} \label{HT_scan}
  \caption{Injected HT power scan for argon, chlorine and sulfur versus the He-like M1 intensity (in arbitrary units).}
% Files around???
\end{figure}

During the optimization of the apparatus, we studied the M1 line 
intensity as a function of the injected HF power.
As expected, we observed a strong dependence between the M1 intensity and the HF power (see fig.~\ref{HT_scan}).
In contrast, we noted an unexpected behavior of the maxima of the curves, whose HF intensity does not increase with the ionization energy of the He-like ion.

%For all the injected gases, we observed  a strong threshold effect. 
%As expected,  the HF power threshold value for M1 observation changes from atom to atom, 
%depending on the ionization energy of the element (see fig.~\ref{HT_scan}).

\hfill
\begin{figure}[h]
\includegraphics[height=.3\textheight]{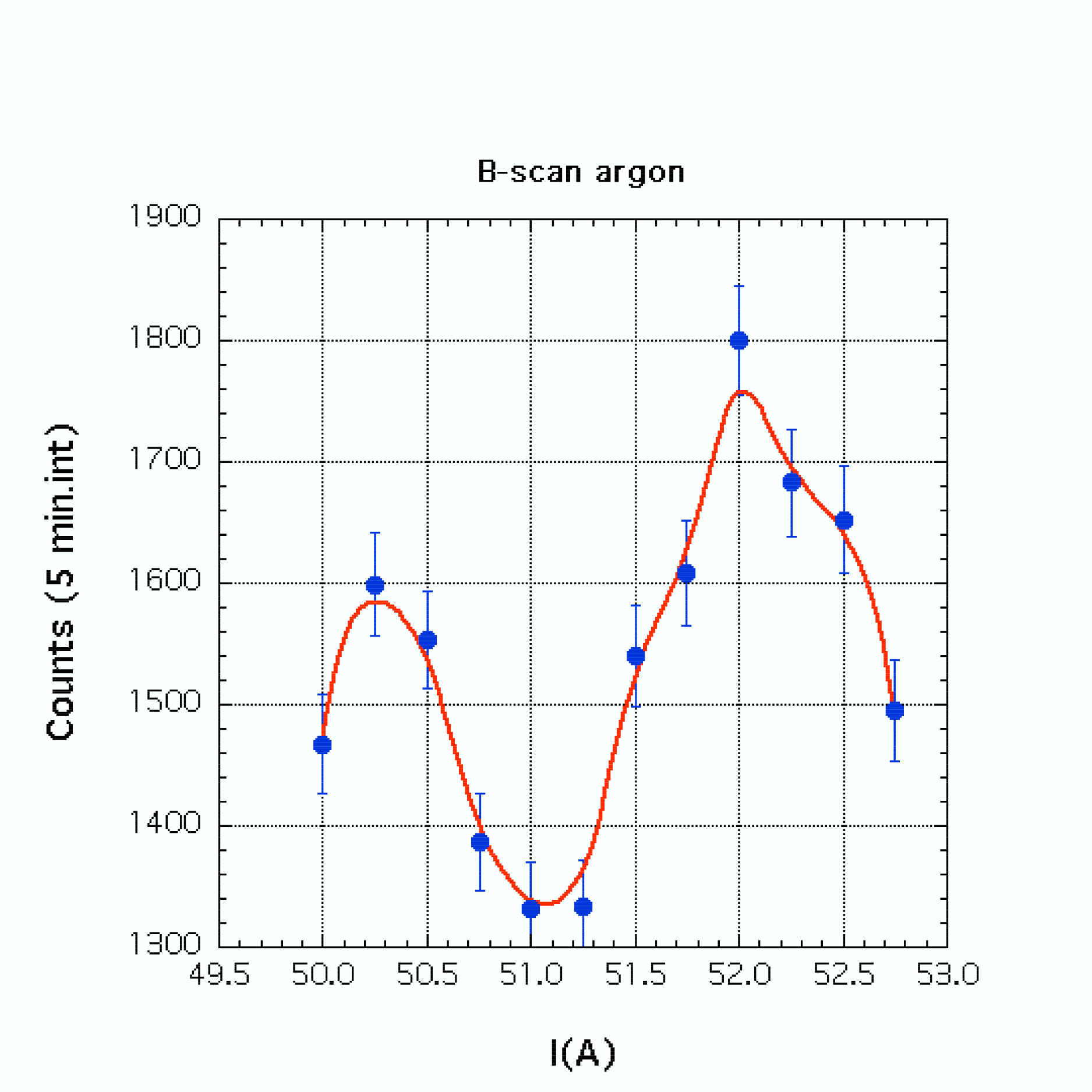} \label{B-scan}
  \caption{He-like argon M1 line intensity dependency versus the superconducting coils
current. 55~A corresponds to a magnetic field value of 2.35~kG at the center of the ECRIT.}
% Files around???
\end{figure}

%Moving at the same time the crystal and the detector it is possible to look at different parts of 
%the plasma. Studying the dependency of the He-like M1 line it is possible measure
%the He-like ions longitudinal distribution.
%With our set-up we could approximately focalize the spectrometer in to the plasma in just 
%few cases because the plasma-crystal distance is fixed at 2200~mm and  the focalization condition
%impose a plasma-crystal distance equal to $R\ \sin(\Theta_B)$, where R is the crystal curvature 
%radius and $\Theta_B$ is the Bragg angle. Moreover the collimator positioned at 150~mm in front of the 
%plasma can influence the target scan at the plasma borders. The best condition to operate a target 
%scan had been obtained using quartz [10-1] crystal with the sulfur spectrum $R\ \sin(\Theta_B) = 2277$. As show in fig.~\ref{target-scan}
%we can observe a small plasma position dependency on the M1 line intensity which decrease slithy on the plasma center.
%A better measurement could be done with a pinhole x-ray camera imaging technique recently developed~\cite{Biri2004}.
%\begin{figure}
%\includegraphics[height=.3\textheight]{target-scan-sulfur-Qz101} \label{target-scan}
% Files around 504
%  \caption{Target scan of the ECR plasma with sulfur using quartz [10-1] crystal}
%\end{figure}

During the ECRIT parameters adjustment, we observed a non-trivial dependency
of the M1 intensity against the longitudinal magnetic field.
As shown in fig.~\ref{B-scan}, we observe the presence of two distinct maxima in the
M1 intensity-coil current relationship.

\subsection{Atomic Spectra}
One of the most important goals of the spring 2004 run was the high-precision measurement of the 
X-ray spectra of argon, chlorine and sulfur. With 1--2 hours  maximum
acquisition time, we obtained high-statistics spectra of {He-}, Li- and Be-like ionic states
of these elements.
A crystal spectrometer like ours can only measure energy differences between atomic transitions. A reference is thus needed.
Due to the lack of high quality reference line in neutral atom X-ray spectra (see, e.g., \cite{agis2003}) we used as a reference He-like 
$1s2s\;^{3}S_1 \to 1s^2\;^{1}S_0\; M1$ transition. All the transition energies provided in the present work are based on the M1 theoretical transition energy calculated with a multi-configuration Dirac-Fock code~\cite{Costa2001,MCDF}.
The peaks in the spectra were fitted with a simulated spectrometer response function which was 
convoluted to a Gaussian.
The response function was obtained through a Monte Carlo X-ray tracking simulation based on the
theoretical reflection function of the crystal obtained with the XOP code~\cite{XOP}.
The reliability of the simulation had previously been tested during the crystal response 
function study~\cite{Simons2004}.
The results obtained in highly-charged argon and sulfur spectroscopy have an
unprecedented precision of the order of 10~meV and they agree with the previous 
experimental values and theoretical predictions (see tables~\ref{argon_en},~\ref{sulfur_en}).

%The spectrometer we used for the experience can measure only relative energies, i.e.,
%it need a calibration line. 
%For this propose we choose as reference the He-like M1 line using as the theoretical
%transition energy~\cite{Costa2001}.
%As we can observe in figure \ref{argon_tot} we could observe the
%all the principal He-like argon transition to the fundamental level. 

\begin{figure}[h]

\includegraphics[width=.45\textwidth]{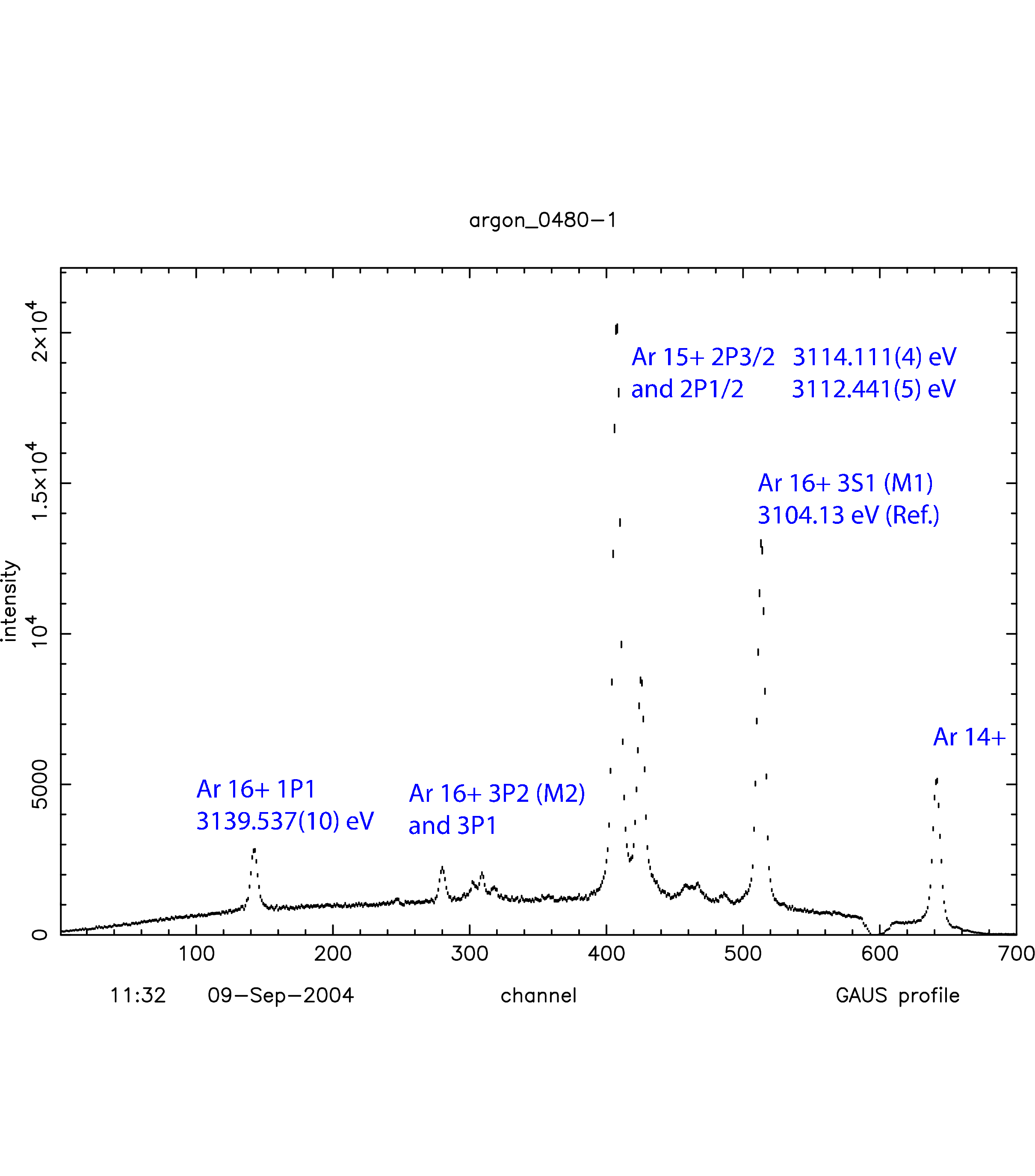} \label{argon_tot}
  \caption{He-like and Li-like argon spectrum using quartz [10-1] crystal. 
HF power injected $P = 400\ W$. argon partial pressure $p_{Ar}= 5 \cdot 10^{-9}mbar$, 
total pressure $p_{tot}= 5 \cdot 10^{-7}mbar$}
\end{figure}
\hfill
\begin{figure}[h]

\includegraphics[width=.45\textwidth]{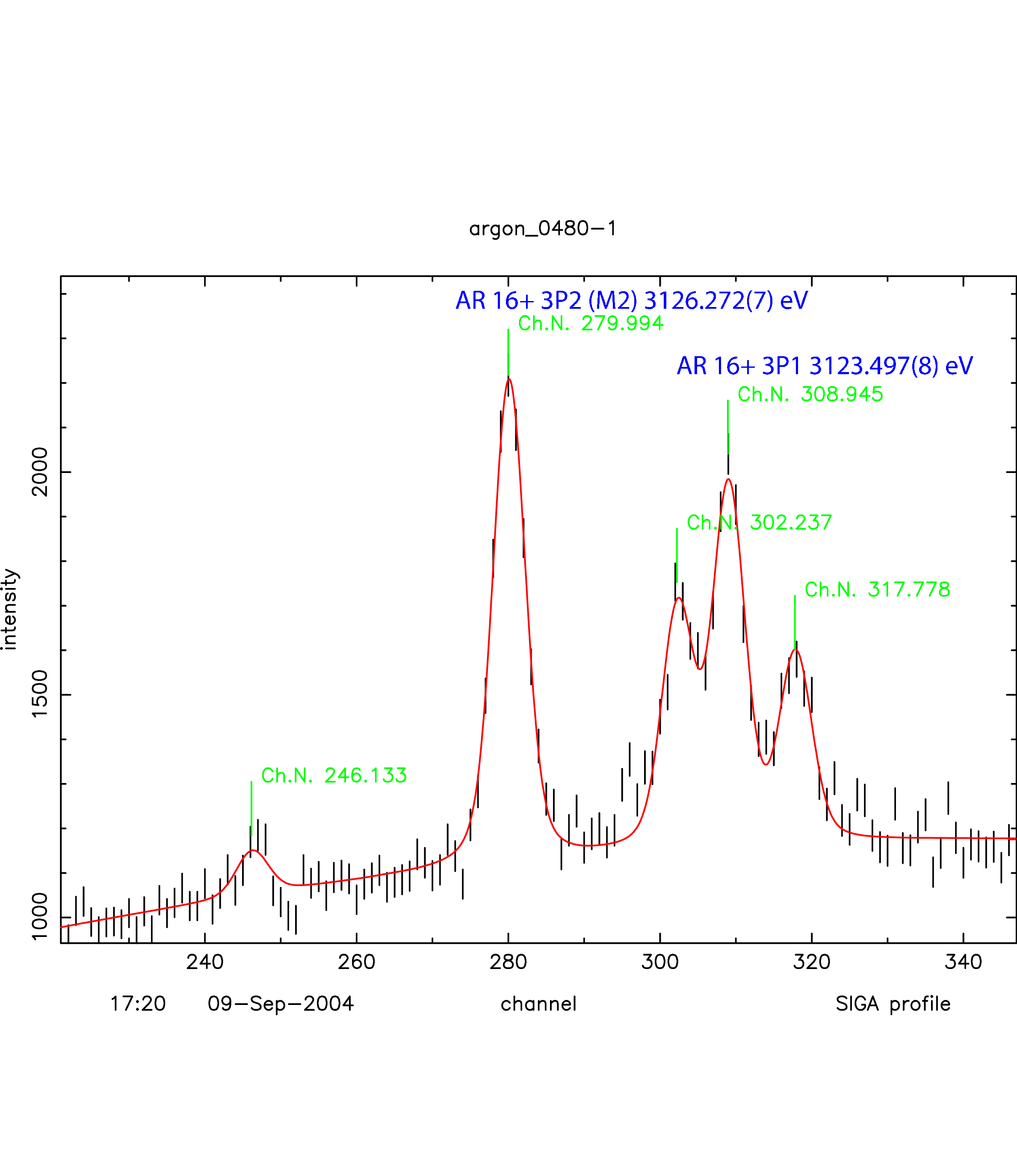} 
  \caption{Detail of the spectrum in figure \ref{argon_tot} around the $1s2p\;^{3}P_n \to 1s^2\;^{1}S_0$ transitions}
\end{figure}
% As a peak shape for the fit of the spectra, we used 
%the simulated response function of the spectrometer convoluted with a Gaussian shape.

%In figure \ref{argon_tot} we can observe the typical spectrum we obtained for argon 
%using quartz [10-1] crystal.
%The fit results are presented in table \ref{argon_en} with the comparison to the 
%theoretical values and the precedent experiments values.

\begin{table}[h]
\begin{tabular}{llll}
\hline
& & \tablehead{1}{r}{b}{$1s2p\;^{1}P_1 \to$\\$\to 1s^2\;^{1}S_0$}
  & \tablehead{1}{r}{b}{$1s2p\;^{1}P_1 \to$\\$\to 1s2p\;^{3}P_1$}\\
\hline
  & {Costa~\cite{Costa2001} (th.)}              & 3139.57      & 16.05\\
  & {Plante~\cite{Plante1994} (th.)}            & 3139.6236    & 16.0484\\
  & {Lindgren~\cite{Lindgren2001} (th.)}       &              & 16.048\\
  & {Deslattes~\cite{Deslattes1984} (exp.)} & 3139.553(36) & 16.031(72)\\
  & {This Work}               & 3139.537(10) & 16.040(17)\\
  & (prelminary results)\\

\hline
\end{tabular}
\caption{He-like argon energy transitions in eV}
\label{argon_en}
\end{table}

\begin{table}[h]
\begin{tabular}{llll}
\hline
& & \tablehead{1}{r}{b}{$1s2p\;^{1}P_1 \to$ \\$\to 1s^2\;^{1}S_0$}
  & \tablehead{1}{r}{b}{$1s2p\;^{1}P_1 \to$\\$\to 1s2p\;^{3}P_1$}\\
\hline
  & MCDF~\cite{MCDF} (th.)                       & 2460.6169  & 13.4875\\
  & Plante~\cite{Plante1994} (th.)                & 2460.6707  & 13.4857\\
  & Schleinkofer~\cite{Schleinkofer1982} (exp.)& 2460.67(9) & 13.62(20)\\
  & This Work                    & 2460.608(9)& 13.483(16)\\
  & (prelminary results)\\

\hline
\end{tabular}
\caption{He-like sulfur energy transitions in eV}
\label{sulfur_en}
\end{table}

\section{Outlook and Conclusion}
In this paper, we demonstrated once more the possibility of using high-precision X-ray spectroscopy with a Bragg spectrometer in the
study of ECR ion sources. Moreover, we presented new results on X-ray 
transition energies in highly-charged argon, chlorine and sulfur.

The next steps will consist in finishing the data analysis of lower-charge states (Li- and Be-like) of argon and sulfur,
in analyzing  the chlorine spectra 
and in studying the injected HF power dependency of the satellite transition in the He-like ions spectra.

%%%%%%%%%%%%%%%%%%%%%%%%%%%%%%%%%%%%%%%%%%%%%%%%
%% BACKMATTER
%%%%%%%%%%%%%%%%%%%%%%%%%%%%%%%%%%%%%%%%%%%%%%%%

\begin{theacknowledgments}
The technical support of L.~Stohwasser, H.~Schneider and D.~St\"uckler was essential
in obtaining the results described here. The advice and help of D.~Hitz and K.~Stiebing during the 
preparation of the ECRIT experiment is warmly acknowledged.
\end{theacknowledgments}

%% If the bibliography is produced without BibTeX comment out the

\bibliographystyle{aipproc}   % if natbib is available

\bibliography{ecris04_trassinelli}

\hyphenation{Post-Script Sprin-ger}
\begin{thebibliography}{17}
\expandafter\ifx\csname natexlab\endcsname\relax\def\natexlab#1{#1}\fi
\providecommand{\enquote}[1]{``#1''}
\expandafter\ifx\csname url\endcsname\relax
  \def\url#1{\texttt{#1}}\fi
\expandafter\ifx\csname urlprefix\endcsname\relax\def\urlprefix{URL }\fi

\bibitem[Gotta(2004)]{Gotta2004}
Gotta, D., \emph{Prog. Part. Nucl. Phys.}, \textbf{52}, 133--195 (2004).

\bibitem[Biri et~al.(2000)]{Biri2000}
Biri, S., Simons, L., and Hitz, D., \emph{Rev. Sci. Instrum.}, \textbf{71},
  1116--18 (2000).

\bibitem[Anagnostopoulos et~al.(2003{\natexlab{a}})]{Simons2003}
Anagnostopoulos, D.~F., Biri, S., Boisbourdain, V., Demeter, M., Borchert, G.,
  Egger, J.~P., Fuhrmann, H., Gotta, D., Gruber, A., Hennebach, M., Indelicato,
  P., Liu, Y.~W., Manil, B., Markushin, V.~E., Marton, H., Nelms, N., Rusi
  El~Hassanii, A., Simons, L.~M., Stingelin, L., Wasser, A., Wells, A., and
  Zmeskal, J., \emph{Nucl. Instrum. Methods B}, \textbf{205}, 9--14
  (2003{\natexlab{a}}).

\bibitem[{Pionic Hydrogen Collaboration}(1998)]{proposal}
{Pionic Hydrogen Collaboration}, \emph{PSI experiment proposal R-98.01} (1998),
  \urlprefix\url{http://pihydrogen.web.psi.ch}.

\bibitem[Bernard(1996)]{Bernard1996}
Bernard, C., Ph.D. thesis, Universit\'e J. Fourier, Lyon (1996).

\bibitem[Sadeghi et~al.(1991)]{Sadeghi1991}
Sadeghi, N., Nakano, T., Trevor, D.~J., and Gottscho, R.~A., \emph{J. Appl.
  Phys.}, \textbf{70}, 2552 (1991).

\bibitem[Anagnostopoulos et~al.(2004)]{Simons2004}
Anagnostopoulos, D., Biri, S., Fuhrmann, H., Gotta, D., Gruber, A., Indelicato,
  P., Leoni, B., Simons, L.~M., Stingelin, L., Wasser, A., and Zmeskal, J.,
  \emph{eprint: physics/0408081} (2004).

\bibitem[Nelms et~al.(2002)]{Nelms2002}
Nelms, N., Anagnostopoulos, D.~F., Ayranov, O., Borchert, G., Egger, J.~P.,
  Gotta, D., Hennebach, M., Indelicato, P., Leoni, B., Liu, Y.~W., Manil, B.,
  Simons, L.~M., and Wells, A., \emph{Nucl. Instrum. Meth. A}, \textbf{484},
  419--31 (2002).

\bibitem[Trassinelli(to be published, eprint:
  physics/0409066)]{Trassinelli2004}
Trassinelli, M., \enquote{Precision Spectroscopy Of Pionic Atoms: From Pion
  Mass Evauation To Tests Of Chiral Perturbation Theory,} in \emph{DA$\Phi$NE
  2004 proceeding, Frascati Physics Series}, to be published, eprint:
  physics/0409066.

\bibitem[Anagnostopoulos et~al.(2003{\natexlab{b}})]{agis2003}
Anagnostopoulos, D.~F., Gotta, D., Indelicato, P., and Simons, L.~M.,
  \emph{Phys. Rev. Lett.}, \textbf{91}, 240801 (2003{\natexlab{b}}).

\bibitem[Costa et~al.(2001)]{Costa2001}
Costa, A.~M., Martins, M.~C., Parente, F., Santos, J.~P., and Indelicato, P.,
  \emph{At. Data Nucl. Data Tables}, \textbf{79}, 223--39 (2001).

\bibitem[Indelicato(2004)]{MCDF}
Indelicato, P., private communication (2004).

\bibitem[{Sanchez del Rio} and Dejus(1998)]{XOP}
{Sanchez del Rio}, M., and Dejus, J., \enquote{XOP: Recent development,} in
  \emph{SPIE proceedings}, 1998, p. 3448.

\bibitem[Plante et~al.(1994)]{Plante1994}
Plante, D., Johnson, W., and Sapirstein, J., \emph{Phys. Rev. A}, \textbf{49},
  3519--3530 (1994).

\bibitem[Lindgren et~al.(2001)]{Lindgren2001}
Lindgren, I., {\AA}s\'en, B., Salomonson, S., and M{\aa}rtensson-Pendrill,
  A.~M., \emph{Phys. Rev. A}, \textbf{64}, 062505 (5) (2001).

\bibitem[Deslattes et~al.(1984)]{Deslattes1984}
Deslattes, R., Beyer, H., and Folkmann, F., \emph{J. Phys. B: At. Mol. Opt.
  Phys.}, \textbf{17}, L689--L694 (1984).

\bibitem[Schleinkofer et~al.(1982)]{Schleinkofer1982}
Schleinkofer, L., Bell, F., Betz, H., Trolman, G., and Rothermel, J.,
  \emph{Phys. Scr.}, \textbf{25}, 917--923 (1982).

\end{thebibliography}

\IfFileExists{\jobname.bbl}{}
 {\typeout{}
  \typeout{******************************************}
  \typeout{** Please run "bibtex \jobname" to optain}
  \typeout{** the bibliography and then re-run LaTeX}
  \typeout{** twice to fix the references!}
  \typeout{******************************************}
  \typeout{}
 }

\end{document}